\documentstyle[aps]{revtex}

\begin{document}
\author{Tim Bolton}
\address{Kansas State University\\
Manhattan, KS 66506-2601}
\date{August 14, 1997}
\title{Determining the CKM Parameter $V_{cd}$ from $\nu{N}$ Charm Production }
\preprint{KSU-HEP-97-03}
\maketitle

\begin{abstract}
The formalism for extracting the CKM parameter $V_{cd}$ from $\nu N$
production of charm is discussed in some detail. The various model
assumptions needed are clearly pointed out. A direct determination from
neutrino induced dimuon production requires $\nu N$ charm production data, $%
\nu N$ charm hadronization data, and the semi-muonic branching ratios for
charmed hadrons. Hadronization data from FNAL E531 is re-analyzed to take
advantage of better-determined properties of the charmed hadrons. A small
bias in the original published result is removed. Neutrino induced charm
fragmentation is compared to $e^{+}e^{-}$ charm fragmentation functions; the
data are consistent with a common hadronization scheme. An updated value of
the mean semi-muonic branching ratio for charmed hadrons produced in $\nu N$
scattering for $E_\nu >30$ GeV is obtained. This value is used to determine $%
V_{cd}$ and its associated uncertainties. Prospects for improving the $%
V_{cd} $ measurement to test the unitarity limit of the CKM matrix are
described.
\end{abstract}

\section{Introduction}

The CKM parameters $|V_{cd}|$ and $|V_{cs}|$ can presently only be measured
via the neutrino production of charm at high energies. This paper summarizes
the current state of knowledge of these CKM\ parameters and estimates the
possible precision that will be achieved in future experiments. Much of the
material has appeared previously in unpublished form\cite{nevis 1501}, and a
condensed summary of this article will appear in a forthcoming review\cite
{rmp}. This document is intended to update the previous result and to
provide more details for specialists. The timing is propitious as NuTeV,
Nomad, and Chorus should shortly produce new experimental results.

In the standard three generation CKM matrix, unitarity and the precise
determinations of $|V_{ud}|$ and $\left| V_{us}\right| $ tightly constrain $%
|V_{cd}|$ and $|V_{cs}|.$ This can be easily appreciated in the Wolfenstein
parameterization\cite{wolfenstein} of the CKM matrix, in which the mixing
between different generations is parameterized as $\sin \theta _{12}=\lambda 
$, sin$\theta _{23}=$$A\lambda ^2$, and $\sin \theta _{13}e^{i\delta
}=A\lambda ^3(\rho +i\eta )$. In this scheme 
\begin{equation}  \label{wolfenstein-Vcd}
\left| \frac{V_{cd}}{V_{us}}\right| =1+A^2\lambda ^4(\rho -\frac 12)\simeq
1+(2.4\times 10^{-3})A^2(\rho -\frac 12),
\end{equation}
and 
\begin{equation}  \label{wolfenstein-Vcs}
\left| \frac{V_{cs}}{V_{ud}}\right| =1-\frac 12A^2\lambda ^4\simeq
1-(1.2\times 10^{-3})A^2.
\end{equation}
Since $A$ and $\rho $ are known to be of order one from measurements of $%
V_{cb}$ and $V_{ub}$ at CLEO and Argus, $|V_{cd}|$ and $|V_{cs}|$ must be
within a few parts per thousand of $\left| V_{ud}\right| $ and $\left|
V_{us}\right| $.

On the other hand, if three generation unitarity is not assumed the coupling
of $|V_{cd}|$ to $|V_{us}|$ is not as tight. For example, in a four
generation unitary CKM matrix, mixing between the second and fourth
generation ($\sin \theta _{24})$ could allow $\left| V_{us}\right| -\left|
V_{cd}\right| \leq 0.03$ while maintaining $\left| V_{ud}\right| ^2+\left|
V_{us}\right| ^2=1.$ In summary, the standard model predicts $|V_{cd}|$ and $%
|V_{cs}|$ to a level of $\pm 0.1\%$. Any larger deviation would indicate new
physics, and deviations as large as $10\%$ are interesting.

Section 2 introduces the necessary formalism and points out assumptions that
need to be made in the analysis of neutrino experiments and external input
that is necessary. Section 3 summarizes the present state of knowledge of $%
|V_{cd}|$ and $|V_{cs}|$ and the auxiliary quantities that are needed to
extract these parameters. Section 4 considers techniques of improving the
measurements and possible systematic limitations; and estimates the
sensitivity new experiments at Fermilab and CERN.

\section{Neutrino Production of Charm}

\subsection{Leading Order Quark-Parton Formalism}

At a fixed reference 4-momentum transfer $Q_0^2$, the isoscalar cross
section for the neutrino production of charm can be written to lowest order
in QCD as 
\begin{equation}  \label{NuLOcharm}
\begin{array}{c}
{\displaystyle {d\sigma (\nu N\rightarrow \mu ^{-}cX) \over dxdy}}%
=%
{\displaystyle {G_F^2ME \over \pi}}%
\times \\ 
(1- \frac{m_c^2}{2ME\xi })\cdot \Theta (y-\frac{m_c^2}{2ME\xi })\cdot \Theta
(\xi -\frac{m_c^2}{2ME})\cdot \Theta \left[ 2MEy(1-x)+M^2-M_C^2\right] \times
\\ 
\left\{ \left| V_{cd}\right| ^2\left[ u_V(\xi ,Q_0^2)+d_V(\xi
,Q_0^2)+u_S(\xi ,Q_0^2)+d_S(\xi ,Q_0^2)\right] +\left| V_{cs}\right|
^22s(\xi ,Q_0^2)\right\}
\end{array}
,
\end{equation}
and for anti-neutrinos as 
\begin{equation}  \label{NubarLOcharm}
\begin{array}{c}
{\displaystyle {d\sigma (\bar \nu N\rightarrow \mu ^{+}\bar c\bar X) \over dxdy}}%
=%
{\displaystyle {G_F^2ME \over \pi}}%
\times \\ 
(1- \frac{m_c^2}{2ME\xi })\cdot \Theta (y-\frac{m_c^2}{2ME\xi })\cdot \Theta
(\xi -\frac{m_c^2}{2ME})\cdot \Theta \left[ 2MEy(1-x)+M^2-M_C^2\right] \times
\\ 
\left\{ \left| V_{cd}\right| ^2\left[ \bar u(\xi ,Q_0^2)+\bar d(\xi
,Q^2)\right] +\left| V_{cs}\right| ^22\bar s(\xi ,Q_0^2)\right\}
\end{array}
.
\end{equation}
where in the above two expressions:

\begin{itemize}
\item  $G_F$ is the Fermi constant, $M$ is the nucleon mass and $E$ is the
incident neutrino energy.

\item  $m_c$ is an effective charm mass parameter, and $M_C$ is the lowest
mass charmed hadronic system allowed by conservation laws.

\item  $y$ is the fraction of neutrino energy transferred to the hadronic
system, and $\xi $ is the fraction of the proton's momentum carried by
(massless) struck quark. In terms of the Bjorken scaling variable $x=\frac{%
Q^2}{2MEy}$, $\xi =$$\xi (x,Q_0^2)=x(1+\frac{m_c^2}{Q_0^2})$.

\item  $\frac 12\left[ u_V(\xi ,Q_0^2)+d_V(\xi ,Q_0^2)\right] =v(\xi ,Q_0^2)$
is the valence quark momentum distribution of an isoscalar nucleon, and $%
\frac 12\left[ u_S(\xi ,Q_0^2)+d_S(\xi ,Q_0^2)\right] =\bar{q}(\xi ,Q_0^2)$
is the light sea quark distribution. Valence and sea quark distributions are
practically defined via $d_V(\xi ,Q_0^2)=d(\xi ,Q_0^2)-\bar{d}(\xi ,Q_0^2)$
and $d_S(\xi ,Q_0^2)=\bar{d}(\xi ,Q_0^2)$, and similarly for $u$-quarks.

\item  $s(\xi ,Q_0^2)=\bar{s}(\xi ,Q_0^2)$ is the strange quark momentum
distribution inside the proton.
\end{itemize}

The second line in the two cross section formulas above incorporates
threshold effects. The first factor $(1-\frac{m_c^2}{2ME\xi })$ is a
kinematic factor reflecting the spin $\frac 12$ character of the quarks and
leptons. The second factor $\Theta (y-\frac{m_c^2}{2ME\xi })$ imposes the
minimum inelasticity requirement for charm production. The third factor $%
\Theta \left[ \xi -\frac{m_c^2}{2ME}\right] $ imposes the requirement that
the neutrino-struck quark system have sufficient invariant mass to form a
charm quark. The fourth factor $\Theta \left[ 2MEy(1-x)+M^2-M_C{}^2\right] $
forces the invariant mass of the final state hadronic system to exceed the
minimum that is compatible with the presence of a charmed hadron and a
baryon. Real thresholds will not likely be so sharp; and these factors are
probably approximations.

Under a number of assumptions detailed below, the major difference between
neutrino and anti-neutrino production of charm is the possibility of
production off the valence quarks in the neutrino case that is absent in the
anti-neutrino mode. The sensitivity to $\left| V_{cd}\right| $ follows if
one can isolate the valence quark contribution to charm production. This may
be accomplished in principle by either measuring the charm cross section at
high $x$, where sea quark distributions are small; or by subtracting the
anti-neutrino cross section from the neutrino cross section. The CKM element 
$V_{cs}$ always appears in combination with $s(\xi ,Q^2)$, which is not yet
an independently measured quantity.

The model assumptions that must be addressed are divided into ``partonic''
and ``hadronic'' quantities below.

\subsection{Partonic Level Issues in $\nu N$ Charm Production}

\subsubsection{Higher Order QCD Effects}

The most obvious question at the partonic level is that of higher order QCD
effects. These have been addressed by a number of authors\cite{CharmNLO}.
While the form of the cross sections becomes much more complex in detail,
the essential structure remains the same as far as the CKM\ matrix elements
are concerned , i.e., 
\begin{equation}  \label{NuNLOcharm}
\begin{array}{c}
{\displaystyle {d\sigma (\nu N\rightarrow \mu ^{-}cX) \over dxdy}}%
=%
{\displaystyle {G_F^2ME \over \pi}}%
\times \\ 
\left\{ 
\begin{array}{c}
\left| V_{cd}\right| ^2F_V\left[ m_c,\alpha _S(Q^2),v(\xi ,Q_0^2)\right] +
\\ 
\left| V_{cd}\right| ^2F_S\left[ m_c,\alpha _S(Q^2),v(\xi ,Q_0^2),\bar q(\xi
,Q_0^2),s(\xi ,Q_0^2),G(\xi ,Q_0^2)\right] + \\ 
\left| V_{cs}\right| ^2F_{SS}\left[ m_c,\alpha _S(Q^2),v(\xi ,Q_0^2),\bar q%
(\xi ,Q_0^2),s(\xi ,Q_0^2),G(\xi ,Q_0^2)\right]
\end{array}
\right\}
\end{array}
\end{equation}
and 
\begin{equation}  \label{NubarNLOcharm}
\begin{array}{c}
{\displaystyle {d\sigma (\bar \nu N\rightarrow \mu ^{+}\bar c\bar X) \over dxdy}}%
=%
{\displaystyle {G_F^2ME \over \pi}}%
\times \\ 
\left\{ 
\begin{array}{c}
\left| V_{cd}\right| ^2\bar F_S\left[ m_c,\alpha _S(Q^2),v(\xi ,Q_0^2),\bar q%
(\xi ,Q_0^2),s(\xi ,Q_0^2),G(\xi ,Q_0^2)\right] + \\ 
\left| V_{cs}\right| ^2\bar F_{SS}\left[ m_c,\alpha _S(Q^2),v(\xi ,Q_0^2),%
\bar q(\xi ,Q_0^2),s(\xi ,Q_0^2),G(\xi ,Q_0^2)\right]
\end{array}
\right\}
\end{array}
.
\end{equation}
In these expressions, $F_V$, $F_S$, $F_{SS}$, $\bar F_S$, and $\bar F_{SS}$
are calculable functionals of the parton distributions that depend on the
running coupling constant $\alpha _S(Q^2)$, the charm mass $m_c$, and the
parton distribution functions at the reference $Q_0^2$ . The sea and strange
sea functionals $F_S$ and $F_{SS}$ include contributions from the nucleon's
gluon momentum distribution $G(\xi ,Q_0^2)$ via $W$-gluon fusion. The gluon
distribution also affects $F_S$ and $F_{SS}$, but not $F_V$, via the normal
QCD evolution of the structure functions. Higher order QCD processes do not
affect the dominance of the valence quark distributions at high $x$; thus $%
\left| V_{cd}\right| $ is still measurable from the neutrino cross section
alone. As long as $F_S=$ $\bar F_S$ and $F_{SS}=$ $\bar F_{SS}$, the valence
contribution to charm production can still be isolated by subtracting the
anti-neutrino cross section from the neutrino cross section.

These last assumptions seems quite reasonable, however it is not necessary
for $s(\xi ,Q_0^2)=\bar s(\xi ,Q_0^2)$, for example, at each value of $\xi $%
, nor even for $\int_0^1$ $d\xi ^{\prime }s(\xi ^{\prime },Q_0^2)=\int_0^1$ $%
d\xi ^{\prime }\bar s(\xi ^{\prime },Q_0^2)$. The only rigorous requirement
is that $\int_0^1$ $\frac{d\xi ^{\prime }}{\xi ^{\prime }}s(\xi ^{\prime
},Q_0^2)=\int_0^1$ $\frac{d\xi ^{\prime }}{\xi ^{\prime }}\bar s(\xi
^{\prime },Q_0^2)$ since the parton distribution functions used here are
defined as momentum distributions. A mechanism that could generate $s(\xi
,Q_0^2)\neq \bar s(\xi ,Q_0^2)$ would be the presence of a $\Lambda K^{+}$
state in the proton's wave function\cite{FockSpace}. If the neutrino
interacts with the proton while it is in this state, there will be a
manifest asymmetry between the $s$-quark distribution of the $\Lambda $ and
the $\bar s$ quark distribution of the $K^{+}$. It would be difficult to
combine the neutrino and anti-neutrino data together if $s(\xi ,Q_0^2)\neq 
\bar s(\xi ,Q_0^2)$ or $d_S(\xi ,Q_0^2)\neq \bar d(\xi ,Q_0^2)$. All that
can be stated is that all measurements up to now are consistent with these
assumptions.

The most serious theoretical issues in the treatment of higher order terms
in neutrino charm production are probably the questions of scale dependence
and the treatment of mass thresholds in the evolution of the running
coupling constant. Whether these issues affect CKM matrix parameters depends
on experimental design.

\subsubsection{Partonic $p_T$ and Mass Effects}

The parton model of deep inelastic scattering assumes that the struck quarks
are massless objects moving colinear with the proton. Neither of these
assumptions need be true. Quarks and gluons inside the nucleon will carry
transverse momentum, both due to their confinement in the finite volume of
the nucleon (``intrinsic'' $p_T$) and due to hard gluon radiation and $q\bar{%
q}$ pair production. These effects appear in the longitudinal structure
function of the nucleus 
\begin{equation}
R_L(\xi ,Q^2)\equiv \frac{F_2(\xi ,Q^2)}{2xF_1(\xi ,Q^2)}\cdot (1+\frac{%
4M^2\xi ^2}{Q^2}).  \label{Rlong}
\end{equation}
The longitudinal structure function can be calculated perturbatively for
high $Q^2$; at lower values of the momentum transfer, higher twist
contributions may become important. From the point of view of charm
production, the most serious question is the effect of the charm quark mass
threshold on $R_L$.

The charged current charm production cross section contains a longitudinal
piece even at lowest order just due to the charm mass. To see this simply,
consider production off valence quarks, and note that 
\[
1-\frac{m_c^2}{2ME\xi }\simeq 1-\frac{m_c^2/Q^2}{1+m_c^2/Q^2}\cdot y. 
\]
Then, 
\[
\frac{d\sigma (\nu d_V\rightarrow \mu ^{-}cX)}{dxdy}\propto v(\xi
,Q^2)\left[ (1-y)+\frac 1{1+m_c^2/Q^2}\cdot \frac{y^2}2+\frac 1{1+m_c^2/Q^2}%
\cdot (y-\frac{y^2}2)\right] . 
\]
The decomposition in $y$ isolates the structure functions, from which it
follows, neglecting the $\frac{4M^2\xi ^2}{Q^2}$ term, that 
\[
R_L^{LO-charm}(\xi ,Q^2)\simeq 1+m_c^2/Q^2. 
\]

While it seems reasonable to consider the $d$-quark to be massless, the
strange quark might be expected to have a current quark mass of $\approx $%
300 MeV$/c^2$. The modification of the kinematics due to this effect has
been treated by Tung {\it et al.}\cite{CharmNLO}{\it , }who have also
considered the effects of the target proton mass. There is little effect on
the differential cross section due to $m_s$ or $M$.

Other mass effects could enter in a dependence of the $M_C^2$ threshold on
the struck quark type. In the case of neutrino-nucleon scattering, the
thresholds are $M_C=M_{\Lambda _C}$, $M_{\Lambda _C}+M_\pi ,$ and $%
M_{\Lambda _C}+M_K$ depending on whether the struck quark is $d$-valence, $d$%
-sea, or $s$, respectively. Antineutrino charm production lacks the valence
channel; and the thresholds are slightly higher for $\bar d$ and $\bar s$
struck quarks than for the corresponding neutrino case: $M_C=M_N+M_D$ for $%
\bar d$ and $M_C=M_N+M_{D_S}$ for $\bar s$.

Both parton mass and $p_T$ effects are small at large $\xi $ where the
valence quark distributions dominate; and corrections to the sea quark
distributions are the same for neutrinos and antineutrinos. On the other
hand, higher twist effects are concentrated at high $\xi $, and are thus of
concern in a $\left| V_{cd}\right| $ extraction.

\subsubsection{Nuclear Effects at the Parton Level}

Most measurements of neutrino charm production are made using nuclear
targets. One must correct for the neutron excess present in the heavier
targets. In addition, there may be more subtle nuclear effects.

A standard assumption made in all deep inelastic scattering analyses is that
strong isospin symmetry is obeyed, i.e., $u^p(x,Q^2)=d^n(x,Q^2)$, and $%
d^p(x,Q^2)=u^n(x,Q^2)$; where the superscripts $p$ and $n$ refer to proton
and neutron, respectively. In addition, it is always assumed that $%
s^p(x,Q^2)=s^n(x,Q^2)$. Recent measurements of the Gottfried sum rule by the
NMC collaboration\cite{GottfriedSumRule} indicate that the stronger
assumption $\bar u(x,Q^2)=\bar d(x,Q^2)$ is not true at small $x$. There are
then two corrections due to the neutron excess in heavier targets, a valence
correction for neutrinos only and a sea correction for neutrinos and
anti-neutrinos: 
\begin{equation}  \label{NuIne0}
\Delta _I^\nu =\frac 12\cdot 
{\displaystyle {(N-Z) \over (N+Z)}}%
\left| V_{cd}\right| ^2\left\{ F_I^V\left[ u_V(\xi ,Q_0^2)-d_V(\xi
,Q_0^2)\right] +F_I^S\left[ \bar u(\xi ,Q_0^2)-\bar d(\xi ,Q_0^2)\right]
\right\} ,
\end{equation}
and 
\begin{equation}  \label{NubarIne0}
\Delta _I^{\bar \nu }=\frac 12\cdot 
{\displaystyle {(N-Z) \over (N+Z)}}%
\left| V_{cd}\right| ^2F_I^S\left[ \bar u(\xi ,Q_0^2)-\bar d(\xi
,Q_0^2)\right] .
\end{equation}

Other nuclear effects are not as obvious; however, so long as all parton
distribution functions used in charm production analysis are extracted from
data using the same or similar nuclear targets, nuclear effects should not
affect the results at the parton level. Nuclear partonic effects can be
important if one must take parton distributions extracted from low atomic
number targets and use them to make corrections in a heavy target. This
happens in the case of the neutron excess correction described above. The
differences between $u_V(\xi ,Q_0^2)$ and $d_V(\xi ,Q_0^2)$ are deduced by
comparing scattering from hydrogen to that of deuterium\cite{F2NvsF2P}. The
results must then be extrapolated to heavier targets by correcting for the
EMC effect. The EMC effect is determined from electroproduction measurements
of the structure $F_2(x,Q^2),$ and not for the valence quark distributions
separately.

\subsection{Hadronization Effects}

\subsubsection{Charmed Quark Fragmentation}

Experiments can only measure charmed hadron production, not charmed quark
production. Some experiments, moreover, do not even detect the charmed
hadrons; instead only the lepton (usually muon) from semi-leptonic charm
decay is measured. Assuming factorization, the charmed hadron $(C)$ cross
section can be connected to the charmed quark $(c)$ cross section via
fragmentation functions: 
\begin{equation}  \label{FragFun}
\begin{array}{c}
{\displaystyle {d\sigma (\nu N\rightarrow \mu ^{-}CX) \over dxdydzdp_T^2}}%
=%
{\displaystyle {d\sigma (\nu N\rightarrow \mu ^{-}cX) \over dxdy}}%
\cdot 
\mathop{\displaystyle \sum }%
\limits_hf_h\cdot D_c^h(z,p_T^2), \\ 
{\displaystyle {d\sigma (\bar \nu N\rightarrow \mu ^{+}\bar CX) \over dxdydzdp_T^2}}%
=%
{\displaystyle {d\sigma (\bar \nu N\rightarrow \mu ^{+}\bar cX) \over dxdy}}%
\cdot 
\mathop{\displaystyle \sum }%
\limits_h\bar f_h\cdot \bar D_{\bar c}^h(z,p_T^2)
\end{array}
.
\end{equation}
Here, $D_c^h(z,p_T^2)$ is the probability distribution for the charmed quark
fragmenting into a charmed hadron of type $h$ carrying a fraction of the
quark's longitudinal momentum $z$ and transverse momentum $p_T$ with respect
to the quark direction. The number $f_h$ is the mean multiplicity of the
hadron $h$ in neutrino production of charm. The analogous objects for
anti-neutrinos are indicated by the barred quantities. Since only one $c$%
-quark is produced in a charged current interaction, one can set the
normalization conditions as 
\begin{equation}  \label{FragNorm}
\displaystyle \int %
_0^1dz\int_0^\infty dp_T^2D_c^h(z,p_T^2)=%
\displaystyle \int %
_0^1dz\int_0^\infty dp_T^2\bar D_{\bar c}^{\bar h}(z,p_T^2)=1,
\end{equation}
and 
\begin{equation}  \label{FracNorm}
\sum\limits_hf_h=\sum\limits_h\bar f_h=1.
\end{equation}

\subsubsection{Neutrino vs. Anti-neutrino Hadronization}

It need not be the case that $f_h=\bar f_h$ or $D_c^h(z,p_T^2)=\bar D_{\bar c%
}^{\bar h}(z,p_T^2)$ since the remnant nucleon will not be the same in $\nu
N $ and $\bar \nu N$ scattering at low energy. The threshold behavior, for
example, differs: 
\[
\nu N\rightarrow \mu ^{-}\Lambda _C,\mu ^{-}\Sigma _C; 
\]
but 
\[
\bar \nu N\not{\rightarrow }\mu ^{+}\bar \Lambda _C,\mu ^{+}\bar \Sigma _C. 
\]
At sufficiently high energies, one expects the charm quark hadronization to
become independent of the remnant nucleon, and consequently $%
D_c^h(z,p_T^2)\rightarrow \bar D_{\bar c}^{\bar h}(z,p_T^2)$ and $%
f_h\rightarrow \bar f_h$

\subsubsection{Quasi-elastic Charm Production}

The cross section for $\nu n\rightarrow \mu ^{-}\Lambda _C,\mu ^{-}\Sigma _C$
were first calculated by Shrock and Lee\cite{ShrockLee}. Using SU(4) flavor
symmetry to relate $\sigma (\nu n\rightarrow \mu ^{-}\Lambda _C)$ to $\sigma
(\nu n\rightarrow \mu ^{-}p)$ , these authors calculated a cross section of $%
\sigma (\nu n\rightarrow \mu ^{-}\Lambda _C)=2.4$ fb for $E_\nu >20$ GeV.
The Fermilab E531 experiment\cite{E531Production} observed three events
consistent with quasi-elastic charm production, from which they obtained a
cross section of $(0.37_{-0.23}^{+0.37})$ fb. The overestimate of the
theoretical calculation is likely due to the substitution of charmed vector
meson masses for the $\rho $ and $A_1$ masses appearing in the $\nu
n\rightarrow \mu ^{-}p$ cross section. Converting the E531 data into the
upper limit $\sigma (\nu n\rightarrow \mu ^{-}\Lambda _C)<1.0$ fb at 90$\%$
C.L., and assuming that the cross section is independent of energy, one
finds that the fraction of neutrino charm production that appears in the
quasi-elastic channel is less than 88$\%$, 24$\%$, 13$\%$, 6$\%$, and 2.5$\%$
for $E_\nu =$10, 20, 30, 50, and 100 GeV, respectively. These numbers allow
one to judge the extent to which the cross section factorizes into
independent production and hadronization terms. Factorization is clearly not
valid for $E_\nu <20$ GeV.

Other non-quark-parton model calculations of neutrino charm production
exist. Einhorn and Lee\cite{CharmVDM} considered a generalization of vector
and axial vector meson dominance (VDM) to the charm sector in the same
spirit as Shrock and Lee's $\Lambda _C$ calculation. The VDM contribution to
the neutrino charm cross section calculated in this model is very large, due
again due the replacement of light vector meson masses with charmed vector
masses in the relevant form factors. The agreement of the data with the
quark parton model of charm production at high energies and the failure of
the VDM to describe the non-charm cross section indicates that VDM is not a
dominant contribution at high energies. For $E_\nu <20$ GeV, this mechanism
could well be important.

\subsubsection{Neutrino Induced Di-lepton Production}

Emulsion experiments can, in principle,  identify all charmed hadrons via
their finite decay length. In practice, there will be efficiency variations
according to the charged multiplicity of the modes; and acceptance for low $z
$ will be limited, particularly at low vales of $E_\nu $ or $y$.

Other experiments that detect only leptons must de-convolve another layer
from the decay process: 
\begin{equation}
\frac{d\sigma (\nu N\rightarrow \mu ^{-}\mu ^{+}X)}{dxdydzdp_T^2dk^{*}d\cos
\theta ^{*}}=\frac{d\sigma (\nu N\rightarrow \mu ^{-}cX)}{dxdy}\cdot 
\mathop{\displaystyle \sum }%
\limits_hf_h\cdot D_c^h(z,p_T^2)\otimes B_\mu ^h\cdot \Gamma _\mu ^h\text{(}%
k^{*}\text{,}\cos \theta ^{*}\text{)},  \label{DimuDist}
\end{equation}
with $B_\mu ^h$ the semi-leptonic branching fraction for the charmed hadron $%
h$, and $\Gamma _\mu ^h$($k^{*}$,$\cos \theta ^{*}$) the joint distribution
function for the muon momentum in the hadron rest frame $k^{*}$ and the
decay angle of the muon with respect to the hadron direction in the hadron
rest frame. The `$\otimes $' symbol indicates that a boost must be performed
along the charmed hadron direction to get to the hadron rest frame. To go
from a dimuon measurement to a charm production measurement thus requires
knowledge of the relative production of the various charmed hadrons, their
fragmentation functions, their semi-leptonic branching fraction, and their
decay distributions.

\ It is possible that the nuclear environment could modify the charmed quark
fragmentation functions for production off heavy targets relative to those
measured in $e^{+}e^{-}$ experiments. Recent measurements\cite{AtomicCharm}
of $D$ meson production using hadron beams on nuclear targets are consistent
with an $A^{1.0}$ behavior of the total production rate and of the $x_F$ and 
$p_T$ distributions. This is evidence against the presence of nuclear
effects in either the production or fragmentation of charmed quarks.

Because of the fragmentation and decay parameters present, neutrino
di-lepton experiments can only measure the quantity $\bar B_\mu \cdot
|V_{cd}|^2$, where the mean semi-muonic branching ratio, defined via 
\begin{equation}  \label{MeanBR}
\bar B_\mu =\sum_hf_h\cdot B{_\mu ^h,}
\end{equation}
is assumed to be the same for neutrinos and antineutrinos. A calculation of $%
\bar B_\mu $ requires knowledge of both the production fractions $f_h$ and
the semi-muonic branching fractions $B_\mu ^h$. In addition, the
fragmentation functions $D_c^h(z,p_T^2)$ and the decay distributions must be
available in order to correct for acceptance and smearing. In practice, a
mean fragmentation function to all charmed hadrons is used for this task.

\subsection{Methods for Extracting $|V_{cd}|$ and $|V_{cs}|$}

The most complete analyses for the extraction of the CKM\ parameters in
neutrino scattering are those of the CCFR collaboration. Distributions of
``visible'' versions of the kinematic variables $E,x,$ and $z$ described
above are fitted to a model with four parameters:

\begin{itemize}
\item  $m_{c,}$ the charm mass;

\item  $\alpha $, a parameter that describes the difference in shape between 
$s(\xi ,Q^2)$ and $\bar{q}(\xi ,Q^2)$;

\item  $\kappa $, the ratio of the nucleon's momentum carried by $s$ quarks
relative to that carried by $u$ and $d$ sea quarks, defined as 
\begin{equation}
\kappa \equiv \frac{\int_0^1s(\xi )d\xi }{\int_0^1\bar{q}(\xi )d\xi }.
\label{KappaDef}
\end{equation}

\item  $\bar{B}_\mu $, the mean semi-muonic branching ratio for charmed
hadrons produced in neutrino anti-neutrino scattering.
\end{itemize}

The two analyses fit the same data to the cross section described in
equation \ref{DimuDist} using either leading order\cite{Rabinowitz} or
next-to-leading order\cite{Bazarko} QCD. To extract values for $\kappa $ and 
$\bar B_\mu $, $\left| V_{cd}\right| $ and $\left| V_{cs}\right| $ are set
to values implied by three generation CKM unitarity. If the unitarity
constraints are dropped, then the CCFR fit parameters change from $\bar B%
_\mu \rightarrow \bar B_\mu \cdot |V_{cd}|^2$ and $\kappa \rightarrow $$\bar %
B_\mu \cdot \frac \kappa {\kappa +2}\left| V_{cs}\right| ^2$.

The CCFR\ analyses use the maximum information in each dimuon event.
However, the procedure couples the desired CKM matrix elements to the
properties of the strange sea. Furthermore, the fits to absolute rates are
sensitive to QCD scale uncertainties.

A cleaner way of obtaining $\left| V_{cd}\right| $ from data is to measure
the ratio of differences\cite{CDHScharm} 
\begin{equation}
R_{\mu \mu }^{-}=\frac{\sigma (\nu N\rightarrow \mu ^{-}\mu ^{+}X)-\sigma (%
\bar{\nu}N\rightarrow \mu ^{+}\mu ^{-}X)}{\sigma (\nu N\rightarrow \mu
^{-}X)-\sigma (\bar{\nu}N\rightarrow \mu ^{+}X)}.  \label{R2MuMinusDef}
\end{equation}
This can be recast in a more useful form for experiments: 
\begin{equation}
R_{\mu \mu }^{-}=\frac{r_{\mu \mu }-r\cdot \bar{r}_{\mu \mu }}{1-r},
\label{R2MuMinusExp}
\end{equation}
where $r_{\mu \mu }=\frac{\sigma (\nu N\rightarrow \mu ^{-}\mu ^{+}X)}{%
\sigma (\nu N\rightarrow \mu ^{-}X)}$ and $\bar{r}_{\mu \mu }=\frac{\sigma (%
\bar{\nu}N\rightarrow \mu ^{+}\mu ^{-}X)}{\sigma (\bar{\nu}N\rightarrow \mu
^{+}X)}$ are the normalized dimuon rates, and $r=\frac{\sigma (\bar{\nu}%
N\rightarrow \mu ^{+}X)}{\sigma (\nu N\rightarrow \mu ^{-}X)}$ is the
well-measured $\bar{\nu}/\nu $ charged current cross section ratio.

To leading order in QCD, 
\begin{equation}  \label{R2MuMinusTheory}
R_{\mu \mu }^{-}=\frac 32\bar B_\mu |V_{cd}|^2\frac{K[m_c,E,v(\xi ,Q^2)]}{1+%
\frac 32|V_{cd}|^2K[m_c,E,v(\xi ,Q^2)]},
\end{equation}
where 
\begin{equation}  \label{KfactorDef}
\begin{array}{c}
K[m_c,E,v(\xi ,Q^2)]= \\ 
{\displaystyle { d\xi dy(1-\frac{m_c^2}{2ME\xi })\Theta (y-\frac{m_c^2}{2ME\xi })\Theta (\xi -\frac{m_c^2}{2ME})\Theta [2MEy(1-x)+M^2-M_C^2]v(\xi ,Q^2) \over \int d\xi v(\xi ,Q^2)}}%
.
\end{array}
\end{equation}
The actual experimental procedure is to measure $R_{\mu \mu }^{-}$ as a
function of neutrino energy. A fit can then be performed to the data with $%
\bar B_\mu |V_{cd}|^2$ and $m_c$ the free parameters. One still must get the
mean branching ratio from other data to obtain the CKM matrix element.

In an emulsion experiment, one could improve the situation by measuring
directly the ratio 
\begin{equation}  \label{RchMinusDef}
R_c^{-}\equiv \frac{\sigma (\nu N\rightarrow \mu ^{-}cX)-\sigma (\bar \nu
N\rightarrow \mu ^{+}\bar cX)}{\sigma (\nu N\rightarrow \mu ^{-}X)-\sigma (%
\bar \nu N\rightarrow \mu ^{+}X)}.
\end{equation}
Again, this can be expressed as a ratio of differences 
\begin{equation}  \label{RchMinusExp}
R_c^{-}=\frac{r_c-r\cdot \bar r}{1-r},
\end{equation}
with $r_c=\frac{\sigma (\nu N\rightarrow \mu ^{-}cX)}{\sigma (\nu
N\rightarrow \mu ^{-}X)}$ and $\bar r_c=\frac{\sigma (\bar \nu N\rightarrow
\mu ^{+}\bar cX)}{\sigma (\bar \nu N\rightarrow \mu ^{+}X)}$. The leading
order expression for $R_c^{-}$ is the same as for $R_{\mu \mu }^{-}$ with
the substitution $\bar B_\mu \rightarrow 1$.

The nice property of $R_{\mu \mu }^{-}$ and $R_c^{-}$ is their functional
dependence only on the valence quark momentum distribution. This fact has
three virtues: the valence quark distribution is better measured; the
valence quark contribution to charm production suffers less suppression from
the charm mass; and the valence quark distribution has a theoretical simpler
QCD\ evolution. Also, since $R_{\mu \mu }^{-}$ depends only on ratios of
cross sections, it should be less susceptible to QCD scale errors. The
cleaner systematics is offset by the need for high statistics since $R_{\mu
\mu }^{-}$ is a ratio of differences.

\section{Current State of $|V_{cd}|$ and $|V_{cs}|$}

\subsection{Measurements of $\bar B_\mu \cdot |V_{cd}|^2$}

The most up-to-date extraction of $B_\mu \cdot |V_{cd}|^2$ is from the CCFR
collaboration\cite{Bazarko}. Employing a next-to-leading order QCD
formalism: 
\[
\bar B_\mu \cdot |V_{cd}|^2=(5.34\pm 0.39\pm 0.24_{-0.51}^{+0.25})\times
10^{-3}{\rm {\ (CCFR-NLO)}.} 
\]
The first two errors are the experimental statistical and systematic errors,
respectively, the systematic error being dominated by the uncertainty in
modeling the charmed quark fragmentation into charmed hadrons. The final
error is attributed to the QCD factorization and renormalization scale
uncertainties. This analysis also indicates that there is a negligible
difference in the value of $\bar B_\mu \cdot |V_{cd}|^2$ if the analysis is
performed to leading order or next-to-leading order in QCD. Accordingly, one
may also use the older result from the CDHS collaboration\cite{CDHScharm}: 
\[
\bar B_\mu \cdot |V_{cd}|^2=(4.1\pm 0.7_{-0.39}^{+0.19})\times 10^{-3}{\rm {%
\ (CDHS-LO)}.} 
\]
The first error is the total experimental error. The second error is the QCD
scale error, which is not given by the original analysis, but is instead
assumed to be the same as in the CCFR measurement.

Combining the two results, assuming that all of the experimental errors are
completely uncorrelated, but that the QCD scale error is totally correlated
and equal to the CCFR value, yields 
\[
\bar{B}_\mu \cdot |V_{cd}|^2=(5.02_{-0.69}^{+0.50})\times 10^{-3}{\rm {\
(CCFR/CDHS)}.} 
\]

The CKM\ element $V_{cs}$ is contained in the CCFR measurement 
\[
\bar B_\mu \cdot \frac \kappa {\kappa +2}\left| V_{cs}\right| ^2=(2.00\pm
0.10\pm 0.06_{-0.14}^{+0.06})\times 10^{-3}, 
\]
where the uncertainties represent statistics, experimental systematics, and
the QCD scale, respectively.

\subsection{Charm Production Fractions}

The charm fractions $f_h$ have only been measured directly in one
experiment, FNAL E531\cite{E531Production}. In checking over the E531
result, a bias was detected in the way that they extracted their charmed
hadron production fractions. Their data is re-fitted with the bias removed.
If one believes that fragmentation functions are universal, one can check
the re-fitted E531 measurements against similar fractions measured in $%
e^{+}e^{-}$ experiments with similar kinematics. The results of this check
are also presented.

\subsubsection{Reanalysis\ Of E531 Data}

In E531, charm could be tagged by the presence of a detached secondary
vertex in the emulsion target. Some 122 events were tagged in this way, 119
which are neutrino induced and three induced by anti-neutrinos. The charge
of each charmed hadron was determined by counting the number of prongs
coming from its decay vertex. Under the assumption that the production of
neutral charmed baryons is negligible at E531 energies, all neutral
candidates are unambiguously identified as $D^0$ mesons. (There is one
neutral particle in the sample that had an identified proton; this event is
a candidate for a neutral charmed baryon, most likely the $\Xi _c^0$, a $csd$
state. However, the reconstructed lifetime of the track is very long,
indicating that this event could be background, or that it could have be
kinematically mis-fitted.) Charged charmed particles, on the other hand,
could be one of three possibilities (again neglecting heavier baryons): a $%
D^{+}$, a $D_S^{+}$, or a $\Lambda _C^{+}$. A good fraction of the $\Lambda
_C^{+}$ could be identified by the presence of a proton in the emulsion. The
majority of the remaining charged events could be a $D^{+}$ or $D_S^{+}$
with essentially equal probability. The bias in the published E531 result is
that they resolved this ambiguity by counting all of the ``toss-up'' events
as $D^{+}$ mesons. They accounted for the possibility of a mistake in this
procedure by increasing the error in the charmed fractions. Given the state
of knowledge of charmed mesons at the time, this procedure was not
unreasonable.

\subsubsection{Re-fitting E531 Data}

Since the complete data set from the experiment is readily available in the
thesis of S. Frederikson\cite{frederikson}, one can re-do the analysis to
remove the bias. The procedure is to try to use all the available
information about each event to construct a likelihood function. The
relevant items are the relative kinematic fit probabilities to the $D^{+}$
vs. $D_S^{+}$ and the proper lifetimes (assuming a particular mass
hypothesis). Unfortunately, the kinematics offers essentially no separation
because of the small $D_S^{+},D^{+}$ mass difference. The lifetimes, on the
other hand, are quite different for the two mesons. Accordingly, a
likelihood function is constructed for each event using only the decay
length information.

The form of the probability function for each event with a charged charmed
hadron is taken to be 
\begin{equation}  \label{E531EventProb}
P(n)=\frac{\epsilon \left[ \ell (n)\right] 
\mathop{\displaystyle \sum }%
_iN_iw_i(n)\frac{M_i}{c\tau _iq_i(n)}e^{-\frac{M_i\ell (n)}{c\tau _iq_i(n)}}%
}{%
\mathop{\displaystyle \sum }%
_iN_i}.
\end{equation}
In this expression:

\begin{itemize}
\item  The index $i$ takes on values $D^{+}$, $D_S^{+}$, $\Lambda _C^{+}$.

\item  The $N_i$ are the number of produced hadrons of each species type.
These are the free parameters in the fit.

\item  The measured quantities $\ell (n)$ and $q_i(n)$ are the decay length
and the momentum for the kinematic fit to the hadron $i$ , respectively.

\item  The $M_i$ are the masses of the charmed hadrons. From the 1992 PDG%
\footnote{%
Most of these measurements have now been improved in the 1996 PDG summary;
however the relative change between the 1992 and 1996 PDG is very small
compared to that between the E531 publication and the 1992 values used here.}%
: $M_{D^0}=1.8645\pm 0005$ GeV/$c^2,$ $M_{D^{+}}=1.8693\pm 0.0005$ GeV/$c^2$%
, $M_{D_S^{+}}=1.9688\pm 0.0007$ GeV/$c^2$, and $M_{\Lambda
_C^{+}}=2.2849\pm 0.0006$ GeV/$c^2$.

\item  The $\tau _i$ are the mean lifetimes of the particles. From the 1992
PDG: $\tau _{D^{+}}=(10.66\pm 0.23)\times 10^{-13}$s, $\tau _{D^0}=(4.20\pm
0.08)\times 10^{-13}$s, $\tau _{D_S^{+}}=(4.50\pm 0.28)\times 10^{-13}$s,
and $\tau _{\Lambda _c^{+}}=(1.91\pm 0.14)\times 10^{-13}$s. Note that these
are known much more accurately than when E531 ran.

\item  The $w_i$ are the acceptance weights for each hypothesis and each
event, 
\begin{equation}
w_i=Q_i(n)\cdot \left[ \int \frac{d\ell \cdot M_i}{c\tau _i\cdot q_i}e^{%
\frac{-\ell \cdot M_i}{c\tau _i\cdot q_i}}\cdot \epsilon (\ell )\right] ^{-1}
\label{E531EvWate}
\end{equation}
These weights include the effects of the finite size and resolution of the
emulsion, of a special $p_T$ cut for one prong charm events, and for general
detector acceptance. Frederikson's thesis gives the weight only for one
hypothesis for each event. Weights for the other hypotheses are estimated by
assuming that the dominant difference in the decay weights is due to the
interplay of the lifetime and the minimum resolvable decay distance. If the
minimum resolvable decay distance is $\ell $$_{min}$, then, approximately, 
\begin{equation}
w_i^{-1}\sim {e}^{-\frac{M_id_{min}}{q_ic\tau _i}},  \label{E531LWate}
\end{equation}
so that 
\begin{equation}
\frac{\log (w_i)}{\log (w_j)}=\frac{M_iq_j\tau _j}{M_jq_i\tau _i},
\label{E531WateScale}
\end{equation}
independent of $\ell $$_{min}$. Thus, if the assumption is correct, one can
scale the acceptance weights using the one given weight and the lifetimes,
masses, and momenta. This scaling should take into account more complicated
effects like the $p_T$ cut placed on one prong events. An explicit
calculation using a minimum cut-off of $\ell $$_{min}=15$ $\mu $m and an
analogous maximum cut-off of $\ell $$_{max}=$ 1.5 cm was also tried, with no
change in the results. $Q_i(n)$ is set to 0 if some information about the
event rules out a particular hypothesis. This occurs typically if no good
kinematic fit existed for one of the hypotheses, or if the presence of an
identified proton in the final state required the charmed hadron to be a
baryon. Otherwise $Q_i(n)=1$.
\end{itemize}

From the probability functions for each event, an extended log-likelihood
function is constructed: 
\[
L=-\sum_n\log {P_n}-N_{obs}\log (N_D\bar w_D+N_{D_S}\bar w_{D_S}+N_{\Lambda
_C}\bar w_{\Lambda _C})+(N_D\bar w_D+N_{D_S}\bar w_{D_S}+N_{\Lambda _C}\bar w%
_{\Lambda _C}) 
\]
The second and third terms above are the log of the Poisson probability
function for observing $N_{obs}$ events given the produced events. The bars
over the acceptance weights indicate that these quantities are averaged. The
Poisson term incorporates the finite statistics of the experiment. The
(negative) log-likelihood function is then minimized. The results of fits
are given in the next section.

\subsubsection{Results of Fits}

The fit results are given in Table \ref{fitresults} in the form of charmed
hadron fractions: $f_{D^0},f_{D^{+}},f_{D_S^{+}},$ and $f_{\Lambda _C}$. The
fractions, their errors, and their correlations are obtained from MINUIT.
(As a technical aside, the $N_i$ defined in the previous section are
expressed in terms of the $f_i$ and the total number of produced charmed
hadron events $N_C$. The $f_i$ and $N_C$ are then allowed to vary. The
constraint $\sum f_i=1$ keeps the number of free parameters the same.). The
correlations between the fractions are give in Table \ref{correlations}.

\begin{table}[tbp]
\begin{tabular}{|c|c|c|c|c|}
\hline
{\bf Energy (GeV)} & {\bf $f_{D^0}$} & {\bf $f_{D^+}$} & {\bf $f_{D_S^+}$} & 
{\bf $f_{\Lambda_C^+}$} \\ \hline
5-20 & $0.32\pm0.11$ & $0.05\pm0.06$ & $0.18\pm0.10$ & $0.44\pm0.12$ \\ 
20-40 & $0.50\pm0.08$ & $0.10\pm0.08$ & $0.22\pm0.08$ & $0.18\pm0.07$ \\ 
40-80 & $0.64\pm0.08$ & $0.22\pm0.09$ & $0.09\pm0.08$ & $0.05\pm0.04$ \\ 
$>80$ & $0.60\pm0.11$ & $0.30\pm0.11$ & $0.00\pm0.06$ & $0.09\pm0.08$ \\ 
$>40$ & $0.61\pm0.06$ & $0.27\pm0.03$ & $0.04\pm0.01$ & $0.07\pm0.02$ \\ 
$>30$ & $0.58\pm0.06$ & $0.26\pm0.06$ & $0.07\pm0.05$ & $0.07\pm0.04$ \\ 
$>20$ & $0.56\pm0.05$ & $0.20\pm0.05$ & $0.11\pm0.04$ & $0.11\pm0.04$ \\ 
$>5$ & $0.53\pm0.05$ & $0.16\pm0.04$ & $0.13\pm0.04$ & $0.17\pm0.04$ \\ 
\hline
\end{tabular}
\caption{E531 Re-fitted Production Fraction Results}
\label{fitresults}
\end{table}

\begin{table}[tbp]
\begin{tabular}{|l|l|l|l|l|l|l|}
\hline
{\bf Energy (GeV)} & C($D^0,D^+$) & C($D^0D_S^+$) & C($D^0\Lambda_C^+$) & C($%
D^+D_S^+$) & C($D^+\Lambda_C^+$) & C($D_S^+\Lambda_C^+$) \\ \hline
5-20 & -0.165 & -0.332 & -0.572 & -0.116 & -0.227 & -0.451 \\ 
20-40 & -0.166 & -0.280 & -0.229 & -0.810 & +0.034 & -0.332 \\ 
40-80 & -0.497 & -0.302 & -0.239 & -0.557 & -0.082 & -0.205 \\ 
$>80$ & -0.633 & -0.001 & -0.304 & -0.002 & -0.410 & -0.001 \\ 
$>40$ & -0.619 & -0.166 & -0.288 & -0.382 & -0.250 & -0.150 \\ 
$>30$ & -0.503 & -0.215 & -0.268 & -0.551 & -0.159 & -0.192 \\ 
$>20$ & -0.404 & -0.284 & -0.375 & -0.544 & -0.181 & -0.138 \\ 
$>5$ & -0.344 & -0.329 & -0.433 & -0.466 & -0.184 & -0.208 \\ \hline
\end{tabular}
\caption{Correlations Among Production Fractions}
\label{correlations}
\end{table}

\subsubsection{Theoretical Expectations}

One can make a simple model for what one expects the charm branching
fractions to be: 
\[
\begin{array}{c}
f_{D^0}=(1-\lambda )\cdot (1-\phi )\cdot \frac{(\frac V2B^{00}+\frac V2%
B^{+0}+\frac 12)}{V+1}, \\ 
f_{D^{+}}=(1-\lambda )\cdot (1-\phi )\cdot \frac{(\frac V2B^{0+}+\frac V2%
B^{++}+\frac 12)}{V+1}, \\ 
f_{D_S^{+}}=(1-\lambda )\cdot \phi ; \\ 
f_\Lambda =\lambda .
\end{array}
\]
The parameters above have the following meanings:

\begin{itemize}
\item  $\lambda $ is the fraction of time the $c$-quark fragments into a $%
\Lambda _c^{+}.$

\item  $\phi $ is the relativity probability the $c$-quark fragments into a $%
D_S^{+}$ vs. any charmed meson.

\item  $V$ represents the relative probability of a $c$-quark fragmenting
into a vector $D^{*}$ meson compared to the probability of fragmenting
directly into a pseudoscalar $D$ meson.

\item  The $B^{xx}$ are $D^{*}$ branching ratios. From the 1992 PDG: $%
B^{00}\equiv B(D^{*0}\rightarrow D^0X)=1.00$, $B^{0+}\equiv
B(D^{*0}\rightarrow D^{+}X)=0.00,B^{++}\equiv B(D^{*+}\rightarrow
D^{+}X)=0.45\pm 0.04$, $B^{+0}\equiv B(D^{*+}\rightarrow D^0X)=0.55\mp 0.04$.
\end{itemize}

On the basis of spin counting rules, one expects $V=3$. Studies of non-charm
fragmentation in $e^{+}e^{-}$ experiments indicate that $\phi \approx 0.15$
and $\lambda \approx 0.10$. The model thus predicts $f_{D^0}=0.54$, $%
f_{D^{+}}=0.22$, $f_{F^{+}}=0.13$, and $f_\Lambda =0.10$.

\subsubsection{$e^{+}e^{-}$ DATA}

The fractions $f_i$can be obtained from CLEO\ data\cite{CleoCharmProduction}
if one accepts the hypothesis that fragmentation functions obtained from $%
e^{+}e^{-}$ experiments with $\sqrt{s}$ $=10.55$ GeV can be used to describe
quark fragmentation in neutrino induced interactions with mean final state
hadronic mass $<W>\approx 10$ GeV$/c^2$. CLEO measured the product of
branching ratio times cross section, $\sigma \cdot B$, for charged and
neutral $D$ and $D^{*}$ mesons, $D_S^{+}$ mesons, and $\Lambda _C^{+}$
baryons for $z=\frac{E_{{\rm {charm}}}}{E_{\max }}>0.5$. They then
extrapolated using a tuned Lund fragmentation model to all $z$. Its is
possible to convert their $\sigma \cdot B$ values into cross sections using
up-to-date charmed hadron decay rates from the 1992 PDG. Table \ref{cleodata}
summarizes the CLEO measurements of $\sigma \cdot B$ and the branching ratio
corrected cross sections integrated over all $z$ and for $z>0.5$.

\begin{table}[tbp]
\begin{tabular}{|l|l|l|l|l|l|}
\hline
{\bf Mode} & {\bf $\sigma\cdot{B}$(pb)} & {\bf $\sigma\cdot{B}$ (pb)} & {\bf %
B.R.(\%) } & {\bf $\sigma$ (nb)} & {\bf $\sigma$ (nb)} \\ 
& $z>0.5$ & all $z$ &  & $z>0.5$ & all $z$ \\ \hline
$D^0\rightarrow{K^-\pi ^+}$ & $27.0\pm1.4$ & $52\pm6$ & $3.65\pm0.21$ & $%
0.74\pm0.06$ & $1.42\pm0.19$ \\ 
$D^+\rightarrow{K^-\pi^+\pi^+}$ & $29.3\pm3.0 $ & $47\pm7$ & $8.0\pm0.8$ & $%
0.37\pm0.05$ & $0.59\pm0.11$ \\ 
$D^{*+}\rightarrow\pi^+D^0(K^-\pi^+)$ & $10.9\pm1.0$ & $17\pm2$ & $%
2.00\pm0.18 $ & $0.54\pm0.06$ & $0.85\pm0.13$ \\ 
$D^{*+}\rightarrow\pi^+D^0(K^-\pi^+\pi^+\pi^-)$ & $23.1\pm1.0$ & $33\pm3$ & $%
4.1\pm0.3$ & $0.56\pm0.06$ & $0.80\pm0.12$ \\ 
$D^{*0}\rightarrow\gamma/pi^0D^0(K^-\pi^+)$ & $19.8\pm4.0$ & $30\pm7$ & $%
3.65\pm0.21$ & $0.54\pm0.11$ & $0.82\pm0.19$ \\ 
$D_S^+\rightarrow\phi\pi^+$ & $5.8\pm1.0$ & $7.2\pm2.0$ & $2.8\pm0.5$ & $%
0.21\pm0.05$ & $0.26\pm0.09$ \\ 
$\Lambda_C^+\rightarrow{p}K^-\pi^+$ & $8.6\pm1.0$ & $13.5\pm4.0$ & $%
3.2\pm0.7 $ & $0.27\pm0.07$ & $0.42\pm0.16$ \\ \hline
\end{tabular}
\caption{Charmed Hadron Production Measurements By CLEO}
\label{cleodata}
\end{table}

From the data in Table \ref{cleodata} one can extract the fractions $f_i$ as
measured in $e^{+}e^{-}$ scattering at$\sqrt{s}=10.55$ GeV. These fractions
are summarized in Table \ref{cleofrac}.

\begin{table}[tbp]
\begin{tabular}{|l|l|l|}
\hline
{\bf fraction} & {\bf $z>0.5$} & {\bf all $z$} \\ \hline
$f_{D^0}$ & $0.47\pm0.03$ & $0.53\pm0.05$ \\ 
$f_{D^+}$ & $0.23\pm0.03$ & $0.22\pm0.03$ \\ 
$f_{D_s^+}$ & $0.13\pm0.02$ & $0.10\pm0.06$ \\ 
$f_{\Lambda_c^+}$ & $0.17\pm0.04$ & $0.16\pm0.05$ \\ \hline
\end{tabular}
\caption{Charm Hadron Fractions Measured By CLEO}
\label{cleofrac}
\end{table}

Using the $D^{*}$ and $D$ data together, CLEO also determined 
\[
\frac V{V+1}=0.85\pm 0.11\pm 0.17. 
\]
This value is consistent with the model expectation of 0.75.

Electron-positron data thus supports the simple model presented above and
yields charmed hadron fractions in agreement with the re-fit E531 results,
but not with the published E531 values.

\subsection{Updated Charm Branching Ratios}

\subsubsection{Direct Extraction of $\bar B_\mu $}

The charm branching ratios were measured in the early to mid eighties at
SLAC. The measurements are not precise; one, the $D_S^{+}$ semi-muonic rate,
is essentially not known at all. One can attempt to use a weak form of the
spectator model of charm decays to improve the precision, but the
correlations introduced result in no practical gain.

The semi-muonic branching ratios for most charmed particles have changed
little recently, with one important exception. (Actually, only one
measurement exists for the semi-muonic rate; the remainder are
semi-electronic rates that are converted assuming $e-\mu $ universality.)
All values below come from the 1996 PDG except for the $D^0$, which also
includes a substantially improved number from CLEO\cite{NewCleoBR}, yielding 
$B_\mu (D^0)=(6.75\pm 0.30)\%$. The $D^{+}$ rate is essentially that of the
Mark III collaboration: $B_\mu (D^{+})=(17.2\pm 1.9)\%$. The $D_S^{+}$ rate
is from a single poor measurement by Mark III\cite{DaleDsBR}: $B_\mu
(D_S^{+})=(5.0\pm 5.4)\%$. The $\Lambda _C^{+}$ rate is an ancient
measurement from Mark II (at SPEAR)\cite{LambdacBR}: $B_\mu (\Lambda _C^{+}\
)=(4.5\pm 1.7)\%$.

Using the direct measurements alone and the re-fitted E531 data yields 
\[
\bar{B}_\mu =0.0919\pm 0.0085_{{\rm CF\ \ }}\pm 0.0041_{{\rm BR\ }}, 
\]
where the first error is the contribution due to the charmed hadron species
fractions and the second error is due to the charmed hadron semi-muonic
branching ratios. Charm hadron production fractions now dominate this error.

\subsubsection{Spectator Model Fits to Branching Fractions}

The $D_S^{+}$ and $\Lambda _C^{+}$ branching ratios are poorly measured. One
can try to improve the situation by imposing a weak form of the spectator
model which requires the semi-electronic partial width $\Gamma _e$ be the
same for all charmed particles. The semi-muonic branching fraction can then
be related to the lifetime of a charmed hadron $X_n$ via 
\[
\hat{B}_\mu ^h=\Gamma _e\cdot \tau _h. 
\]
One can use the spectator model to fit for improved values of the branching
ratios by minimizing 
\[
\chi ^2=\sum_h\frac{\left( \hat{B}_\mu ^h-B_\mu ^h\right) ^2}{\sigma _{\bar{B%
}_\mu ^h}^2}+\frac{\left( \frac{\hat{B}_\mu ^h}{\Gamma _e}-\tau _h\right) ^2%
}{\sigma _{\tau _h}^2} 
\]
with respect to improved values of the branching ratios $\hat{B}_\mu ^h$ and
the common semi-electronic width $\Gamma _e$. The results of the fit are:\ $%
\hat{B}_\mu (D^0)=(6.76\pm 0.30)\%$, $\hat{B}_\mu (D^{+})=(17.2\pm 0.8)\%$, $%
\hat{B}_\mu (D_S^{+})=(7.6\pm 0.4)\%$, and $\hat{B}_\mu (\Lambda _C^{+}\
)=(3.4\pm 0.3)\%$. The spectator model fit introduces large correlations
among the fit parameters: $C_{D^0D^{+}}=0.930$, $C_{D^0D_S^{+}}=0.613$, $%
C_{D^0\Lambda _C^{+}}=0.698$, $C_{D^{+}D_S^{+}}=0.598$, $C_{D^{+}\Lambda
_C^{+}}=0.681$, and $C_{D_S^{+}\Lambda _C^{+}}=0.449$. The correlations must
be included in calculating the error on the fitted mean branching ratio $%
\hat{B}_\mu $. The result is $\hat{B}_\mu =0.0930\pm 0.0088$.

\subsection{The CKM Parameters}

\subsubsection{$|V_{cd}|$ from Neutrino Scattering}

Using the updated value for $\bar{B}_\mu $ obtained above: 
\[
|V_{cd}|=0.232_{-0.019}^{+0.017}{\rm {\ (DIRECT).}} 
\]
The CKM parameter $|V_{cd}|$ is thus known to $\pm 9\%$ from direct
measurement. Of this, the contribution from the uncertainty on $B_\mu $ is $%
\pm $$6.6\%$, from the experimental error on $B_\mu \cdot |V_{cd}|^2$ $4.2\%$%
, and from the QCD scale uncertainty $_{-5.3}^{+2.6}\%$. The measurement is
consistent with the value inferred from the assumed unitarity of the CKM\
matrix of 
\[
|V_{cd}|=0.221\pm 0.003.{\rm {\ (UNITARITY)},} 
\]
but the unitarity prediction is clearly not being tested.

\subsubsection{Other Measurements of $|V_{cd}|$}

There are no other direct measurements of $|V_{cd}|$. The closest are the
measurements of 
\[
\left| \frac{V_{cd}}{V_{cs}}\right| ^2\cdot \left| \frac{f_{+}^\pi (0)}{%
f_{+}^K(0)}\right| ^2=\epsilon \cdot \frac{B(D\rightarrow \pi \ell \nu )}{%
B(D\rightarrow K\ell \nu )} 
\]
by Mark III in the neutral $D$ mode\cite{mark3vcd} and CLEO in the charged $%
D $ mode\cite{cleovcd}. In this expression, $B(D\rightarrow \pi \ell \nu )$
and $B(D\rightarrow K\ell \nu )$ are the measured $D_{\ell 3}$ semi-leptonic
branching fractions; $\epsilon $ is a precisely known kinematic factor; and $%
f_{+}^\pi (0)$ and $f_{+}^K(0)$ are the $D_{\ell 3}$ transition form factors
evaluated at zero momentum transfer to the $\ell \nu $ system. The Mark III
measurement is 
\[
\left| \frac{V_{cd}}{V_{cs}}\right| ^2\cdot \left| \frac{f_{+}^\pi (0)}{%
f_{+}^K(0)}\right| ^2=0.057_{-0.015}^{+0.038}\pm 0.005{\rm {(MarkIII)}.} 
\]
The CLEO measurement is 
\[
\left| \frac{V_{cd}}{V_{cs}}\right| ^2\cdot \left| \frac{f_{+}^\pi (0)}{%
f_{+}^K(0)}\right| ^2=0.085\pm 0.027\pm 0.014{\rm {(CLEO)}.} 
\]
Combining the two results yields 
\[
\left| \frac{V_{cd}}{V_{cs}}\right| ^2\cdot \left| \frac{f_{+}^\pi (0)}{%
f_{+}^K(0)}\right| ^2=0.069\pm 0.020{\rm {(COMBINED)}.} 
\]
The ratio $\left| \frac{f_{+}^\pi (0)}{f_{+}^K(0)}\right| $ is model
dependent and ranges from $0.7-1.4$\cite{FplusRefs}$.$ This translates into $%
30\%$ uncertainty in $|$$V_{cd}|$. If the theoretical uncertainty in the
form factor ratio could be reduced to $10\%$, then the error on $|$$V_{cd}|$
from $D$ meson decay would be $\pm 0.034$ or $15\%$. This is less precise
than the neutrino measurement.

\subsubsection{Limit on the Wolfenstein Parameters $A$ and $\rho $}

The quantity $\left| \frac{V_{cd}}{V_{us}}\right| -1$ can be used, via
equation \ref{wolfenstein-Vcd}, to set the constraint 
\[
A^2\cdot \left| \rho -\frac 12\right| <173{\rm {~at~90\%~C.L.}} 
\]
This is obviously not a very exciting limit. There is no evidence for the
third generation from quark mixing measurements involving only the first two
generations.

\subsubsection{Limit on $\left| V_{cs}\right| $}

If one assumes that the strange sea carries no more momentum than the light
quark sea, i.e., $\kappa \leq 1$, then it follows from the CCFR
next-to-leading-order analysis that 
\[
\left| V_{cs}\right| >0.74{\rm {~at~90\%~C.L.}}
\]
If one arbitrarily assumes that $\kappa =0.5\pm 0.5$, then 
\[
\left| V_{cs}\right| =1.04\pm 0.42,
\]
and 
\[
\left| \frac{V_{cd}}{V_{cs}}\right| =0.23\pm 0.09\text{.}
\]
The errors in these latter two quantities is dominated by the uncertainty in 
$\kappa $. Note that the ratio $\left| V_{cd}/V_{cs}\right| $ derived from
neutrino and antineutrino scattering is independent of $\bar{B}_\mu $.

If $\kappa $ can be independently measured to an accuracy of $\pm 25\%$ from
the structure function difference $xF_3^\nu -xF_3^{\bar \nu }$, then the
errors on $\left| V_{cs}\right| $ and $\left| \frac{V_{cd}}{V_{cs}}\right| $
would be reduced to $10\%$.

\section{Future Measurements of $|V_{cd}|$ in $\nu N$ Scattering}

Four high energy neutrino experiments are either now running or are approved
to run in the next six years. While none of the experiments are optimized
for the study of neutrino charm production, all have the potential to
improve the CKM matrix element measurements. The experiments are summarized
in Table \ref{future}. The Nomad\cite{nomad} and Chorus\cite{chorus}
experiments at CERN are designed to search for $\nu _\mu \rightarrow \nu
_\tau $ oscillations. Nomad features a low mass target with very good
tracking and electron identification. This experiment should be able to
detect charm in both di-lepton modes ($\mu \mu $ and $\mu e$). Their
excellent tracking may also allow for the identification of charm via the ``$%
D^{*}\rightarrow D\pi $ trick''. Chorus is a hybrid emulsion spectrometer.
It's major virtue is its ability to reconstruct charm inclusively via the
identification of the charm decay vertex. This feature serves to boost
statistics, and, more importantly, largely eliminates the need to know the
production, fragmentation, and decay properties of the charmed hadrons.
Fermilab E815\cite{e815} uses the E744/770 Lab E neutrino detector. The
experiment is optimized for precision studies of neutral current
interactions. The feature most relevant for charm studies is the new
sign-selected neutrino beam. This will eliminate the $\nu /\bar{\nu}$
confusion in the dimuon channel and permit a cleaner measurement of $\left|
V_{cs}\right| $, assuming that the strange sea is independently known by
then.

The ultimate neutrino charm production experiments are COSMOS (FNAL E803)%
\cite{e803} and TOSCA\cite{tosca} at CERN. Like Chorus, these experiments
are designed for a high sensitivity search for $\nu _\mu \rightarrow \nu
_\tau $ oscillations using hybrid emulsion spectrometers. E803 will have a
factor of twenty higher statistics than Chorus; and its spectrometer will
have three times better resolution. The higher resolution is crucial to
reduce backgrounds, particularly in one-prong decays of charm. E803 might be
able to achieve a resolution of $\sim 2\%$ on $\left| V_{cd}\right| $. This
is estimated by assuming: a sample of 50,000 reconstructed charm events,
which reduces the statistical error to $\pm 0.003$; a $\times 5$ reduction
in the experimental systematic errors due to the elimination of
fragmentation uncertainties and background; a $\times 5$ reduction in the
QCD scale error via the normalization of charm to single muon production
that is possible with higher statistics; and a $\times 10$ reduction in
production fraction and branching ratios achieved by the ability to
inclusively reconstruct charm. The total $\pm 0.004$ error on $\left|
V_{cd}\right| $ will be comparable to that on $\left| V_{us}\right| $; and
one will thus be able to test the unitarity property of the CKM matrix at a
level that is sensitive to new physics.

\begin{table}[tbp]
\begin{center}
\begin{tabular}{|l|l|l|l|l|}
\hline
{\bf Experiment} & {\bf Target} & {\bf Start} & {\bf CC Sample} & {\bf Charm
Sample} \\ \hline
Nomad (CERN) & low mass & 1994 & $1\times10^6$ & $2\times10^4(e\mu,\mu%
\mu,D^*)$ \\ 
Chorus (CERN) & emulsion & 1994 & $3\times10^5$ & $2\times10^4$(inclusive)
\\ 
NuTeV (FNAL) & iron & 1996 & $3\times10^6$ & $2\times10^4(\mu\mu$) \\ 
COSMOS (FNAL) & emulsion & 2001 & $8\times10^6$ & $4\times10^5$ (inclusive) \\ 
\hline
\end{tabular}
\end{center}
\caption{Future Neutrino Experiments. Event samples are rough estimates.}
\label{future}
\end{table}


\begin{references}
\bibitem{nevis 1501}  T. Bolton, ``Determining the CKM Parameter $V_{cd}$
from $\nu N$ Charm Production'', Nevis preprint R1501 (unpublished), 1995.

\bibitem{rmp}  J. M. Conrad, M. H. Shaevitz, and T. Bolton, ``Precision
Measurements with High Energy Neutrino Beams'', e-print archive:
hep-ex/9707015, submitted to Reviews of Modern Physics, July, 1997.

\bibitem{wolfenstein}  L. Wolfenstein, Phys. Rev. Lett., {\bf 51} (1983)
1945.

\bibitem{CharmNLO}  M.A.G. Aivazis, F.I. Olness, and W.-K. Tung, S.M.U.
Preprint S.M.U.-HEP/93-16, (1993).M.A.G. Aivazis, F.I. Olness and W.-K.
Tung, Phys. Rev. Lett. 65 (1990) 2339. M.A.G. Aivazis, J.C. Collins, F.I.
Olness and W.-K. Tung, SMU-HEP/93-17, to be published in Phys. Rev. D 50
(1994). G. Kramer and B. Lampe, Z. Phys. C 54 (1992) 139. J.J. van der Bij
and G.J. van Oldenborgh, Z. Phys. C 51 (1991) 477.

\bibitem{FockSpace}  M. Burkardt and B.J. Warr, Phys. Rev. {\bf D45} (1992)
958.

\bibitem{GottfriedSumRule}  M. Arneodo {\it et al. (}New Muon
Collaboration), Phys. Rev.{\bf \ D50} (1994) 1.

\bibitem{F2NvsF2P}  P. Amaudruz {\it et al. (}New Muon Collaboration), Nucl.
Phys. {\bf B371} (1992) 3.

\bibitem{ShrockLee}  R.E. Shrock and B.W. Lee, Phys. Rev. {\bf D13} (1976)
2539.

\bibitem{CharmVDM}  M. Einhorn and B. Lee, Phys. Rev {\bf D13} (1976) 43.

\bibitem{E531Production}  N. Ushida {\it et al.} (E531 Collaboration), Phys.
Lett. {\bf B206} (1988) 375.

\bibitem{AtomicCharm}  G.A. Alves {\it et al. }(E769 Collaboration) , Phys.
Rev. Lett. {\bf 70} (1993) 722; M.J. Leitch {\it et al.} (E789
Collaboration), Phys. Rev. Lett. {\bf 72} (1994) 2542.

\bibitem{Rabinowitz}  S.A. Rabinowitz {\it et al.} (CCFR Collaboration),
Phys. Rev. Lett. {\bf 70} (1993) 134.

\bibitem{Bazarko}  A. O. Bazarko {\it et al.} (CCFR Collaboration), Z. Phys.
C65 (1995) 189.

\bibitem{CDHScharm}  H. Abromowicz {\it et al.} (CDHS Collaboration), Z.
Phys. {\bf C15 }(1982) 19.

\bibitem{frederikson}  S. G. Frederikson, University of Ottowa Ph.D. Thesis
(unpublished) 1987.

\bibitem{CleoCharmProduction}  D. Bortoletto (CLEO Collaboration), {\it et
al.}, Phys. Rev. {\bf D37 }(1988) 1719.

\bibitem{1992PDG}  Particle Data Group, M. Aguilar-Benitez {\it et al.} ,
Phys. Rev. D45, Part 2 (1992).

\bibitem{NewCleoBR}  Y. Kubota {\it et al.} (CLEO Collaboration), Phys. Rev. 
{\bf D54} (1996) 2994.

\bibitem{Mark3DBR}  Mark III Collaboration, R.M. Baltrusaitus {\it et al.},
Phys. Rev. Lett. {\bf 54} (1985) 1976. {\it Erratum}, Phys. Rev. Lett. {\bf %
55 }(1985) 638.

\bibitem{HRSD0BR}  H. Abachi {\it et al. (}HRS Collaboration), Phys. Lett. 
{\bf B205} (1988) 411.

\bibitem{DaleDsBR}  Z. Bai {\it et al. (}Mark III Collaboration), Phys. Rev.
Lett. {\bf 65} (1990) 686.

\bibitem{LambdacBR}  E. Vella {\it et al.}, Phys. Rev. Lett. {\bf 48} (1982)
1515.

\bibitem{mark3vcd}  J. Adler {\it et al}. (Mark III Collaboration), Phys.
Rev. Lett. {\bf 62} (1989) 1821.

\bibitem{cleovcd}  M.S. Alam {\it et al}. (CLEO\ Collaboration), Phys. Rev.
Lett. {\bf 71 }(1993)1311.

\bibitem{FplusRefs}  N. Isgur et al., Phys. Rev. {\bf D39} (1989) 799; M.
Bauer, B. Stech, and M. Wirbel, Z. Phys. {\bf C34} (1987) 103; C.A.
Dominguez, Phys. Lett. {\bf B207} (1988) 499; T.M. Aliev, A.A. Ovchinnikov,
and V.A. Slobodenyuk, Triest Report No. IC/89/382; M. Crisafulli, et al,
Phys. Lett. {\bf B223} (1989) 90; G.P. Lepage and S.J. Brodsky, Phys. Rev. 
{\bf D22} (1980) 2157.

\bibitem{nomad}  P. Astier {\it et al.} (NOMAD Collaboration), ``Search for
the Oscillation $\nu _\mu \rightarrow \nu _\tau $'', CERN-SPSLC/91-21 (11
March 1991).

\bibitem{chorus}  N. Armenise {\it et al.}(Chorus Collaboration), ``A New
Search for $\nu _\mu \rightarrow \nu _\tau $ Oscillations'', CERN-SPSC/90-42
(15 December 1990); M. deJong et al., ``A New Search for $\nu _\mu
\rightarrow \nu _\tau $ Oscillations'', CERN-PPE/93 (19 July 1993).

\bibitem{e815}  T. Bolton {\it et al. }(NuTeV Collaboration), ``Precision
Measurements of Neutrino Neutral Current Interactions Using a Sign Selected
Beam'', Fermilab-Proposal-P-815 (1990).

\bibitem{e803}  K. Kodama {\it et al. }(COSMOS\ Collaboration),
```Muon-Neutrino to Tau-Neutrino Oscillations'', Fermilab-Proposal-P-803
(1993).

\bibitem{tosca}  A.S. Ayan, et al. (TOSCA Collaboration), `` A High
Sensitivity Short Baseline Experiment to Search for $\nu _\mu \rightarrow
\nu _\tau $ Oscillation'', CERN-SPSC-97-05 (March, 1997).
\end{references}
\end{document}